\documentclass[iop]{emulateapj}

\begin{document}

\title{Predicting $\alpha$ Comae Berenices Time of Eclipse I:\\A 26 Year Binary will Eclipse within 2 weeks of 25 January 2015}

\author{Matthew W.~Muterspaugh\altaffilmark{1, 2} And Gregory W.~Henry\altaffilmark{2}}
\altaffiltext{1}{Department of Mathematical Sciences, College of Engineering, 
Tennessee State University, Boswell Science Hall, Nashville, TN 
37209 }
\altaffiltext{2}{Center of Excellence in Information Systems, Tennessee State University, 3500 John A. Merritt Blvd., Box No.~9501, Nashville, TN 
37209-1561}

\email{matthew1@coe.tsuniv.edu, gregory.w.henry@gmail.com}

\begin{abstract}
The dwarf stars in the 26 year period binary $\alpha$ Com are predicted to 
eclipse each other in early 2015.  Simulations of randomly selected orbital 
parameters based on the most recently published orbital elements and 
uncertainties show the eclipse will happen within 3 days of MJD 57047 (January 
25, 2015; $1 \sigma$; all simulations fall within 2 weeks of this) and will 
have a depth of several tenths of a magnitude.  The eclipse will last between 
28 and 45 hours, depending on the exact orbital parameters.  Observers are 
encouraged to monitor this unique event.
\end{abstract}

\section{Introduction}

As unlikely as it seems, a pair of bright dwarfs stars (V=4.3) orbits so edge-on as seen by earth ($90.054 \pm 0.010^\circ$) that the system is likely to eclipse despite its very long 26 year period.  Unlike other long period eclipsing binaries such as V383 Scorpii \citep{2013A&A...550A..93G} or $\epsilon$ Aurigae \citep[e.g.][]{2010Natur.464..870K}, the likihood of an eclipse is not enhanced by inflated stars or circumstellar material, but rather the incredibly lucky geometry of the orbit itself.  In addition to the favorable inclination, the orbital eccentricity of 0.5 and longitude of periastron passage of $102^\circ$ increase the the probability of eclipse.  \cite{Hart1989} and \cite{hoffleit1996} identified the possibility for eclipses, and the revised orbit of \cite{2010AJ....140.1623M} showed the eclipses to be a $5\sigma$ certainty.  While the eclipse itself is reliably predicted, the exact timing depends on the values of all the orbital elements.  Within the uncertainties to which those parameters have been determined, the predicted time of eclipse could vary by days.  In this research memo we present results of simulations based on the current best orbit solution and the distributions of eclipse timings, durations, and resulting photometric signals.

\section{Methodology}

We created 100,000 random sample sets of binary orbit parameters based on the orbital model from \cite{2010AJ....140.1623M} and repeated in Table \ref{tab::visualOrbits}.  In each set, random values were selected for each orbital element using a Guassian-distributed random number generator with $1\sigma$ width corresponding the the parameter's formal uncertainty and centered at the best-fit value (e.g.~values of the period were selected as $9485.68 + 0.97\times g$ days, where $g$ is a standard normal deviate random number).  The resulting set of parameters was then used to calculate the sky-projected separation of the binary every minute from MJD 57023 to 57083, to ensure all likely eclipse times were included.  For each set, we recorded the time of closest approach, the distance of closest approach, and the duration over which the binary separation was less than 0.7 mas (the approximate diameters of the stars).  If any set failed to produce a minimum separation less than 0.7 mas, it was flagged as non-eclipsing.  However, in 100,000 trials, no such combination was found.

The distance of closest approach is related to the eclipse depth as
$$
\Delta m = 2.5 \log_{10}\left(1 - \frac{\beta-\sin \beta}{2\pi}\right)
$$
where
$$
\beta = 2 \cos^{-1} \frac{b}{d_*}
$$
with $b$ the projected separation between the centers of the stars and $d_*$ the diameter of one star.  For the simplicity of this model, it is assumed the stars have the same size and are equally luminous (approximately correct for this system).

\begin{deluxetable*}{ccccccc}
\tablecolumns{10}
\tablewidth{0pc} 
\tablecaption{Visual Orbit Parameters\label{tab::visualOrbits}}
\tablehead{ 
\colhead{Period (d)} 
& \colhead{${\rm T_\circ}$ (HMJD)} 
& \colhead{Semimajor Axis (arcsec)} 
& \colhead{Eccentricity} 
& \colhead{Inclination (deg)} 
& \colhead{$\omega$ (deg)} 
& \colhead{$\Omega$ (deg)}\\
\colhead{$\sigma_{\rm P}$} 
& \colhead{$\sigma_{\rm T_\circ}$} 
& \colhead{$\sigma_a$} 
& \colhead{$\sigma_e$} 
& \colhead{$\sigma_i$} 
& \colhead{$\sigma_\omega$} 
& \colhead{$\sigma_\Omega$}}\\
\startdata
9485.68  & 47651.8   & 0.66132   & 0.4957   &  90.054 &  101.689 &  12.221  \\
(0.97) &    (2.6)  & (0.00061) & (0.0010) & (0.010) &  (0.059) &  (0.015)
\enddata
\tablecomments{
The model parameters and fit uncertainties for the binary orbit of $\alpha$ Com from \cite{2010AJ....140.1623M}.
}
\end{deluxetable*}

\section{Results}

\begin{figure}[!ht]
\epsscale{1.1}
\plotone{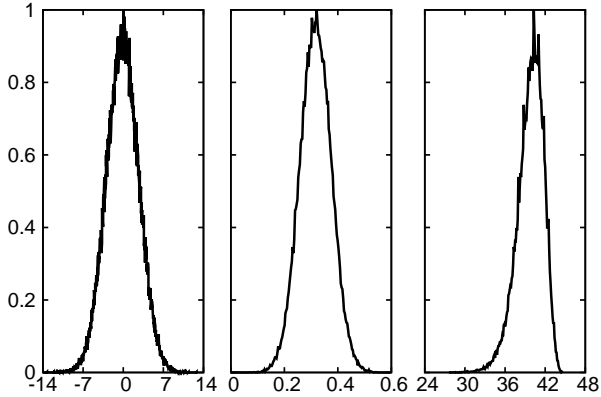}
\caption[HD 114378 ($\alpha$ Com) Eclipse Prediction]
{ \label{fig::114378_eclipse}
(Left) Histogram of the time of the eclipse mid-point, versus days since 25 January 2015 (MJD 57047).  (Middle) Histogram of the closest projected separation of the binary (maximum eclipse) in units of milli-arcseconds.  (Right)  Histogram of the duration of the eclipse from ingress at 0.7 mas separation to egress at the same, in units of hours.
}
\end{figure}

Figure \ref{fig::114378_eclipse} shows histograms of the results from 100,000 simulated orbital parameter sets in terms of the time of eclipse minimum, the minimum projected star separation (maximum eclipse), and the duration of the eclipse.  All 100,000 simulations resulted in eclipses.  The earliest eclipse was at MJD 57033.67 (11 January 2015) and the latest was 57059.76 (6 February 2105).  The closest approach ranged from 0.02 to 0.55 mas.  The eclipse durations fell between 27.9 and 44.6 hours, all longer than 24 hours and enabling observers at all longitudes to observe the event, pending weather (in fact, at declination $+17^\circ$, it is also observable for all latitudes outside the Antarctic Circle, where it is daytime anyways).

\begin{figure}[!ht]
\epsscale{1.0}
\plotone{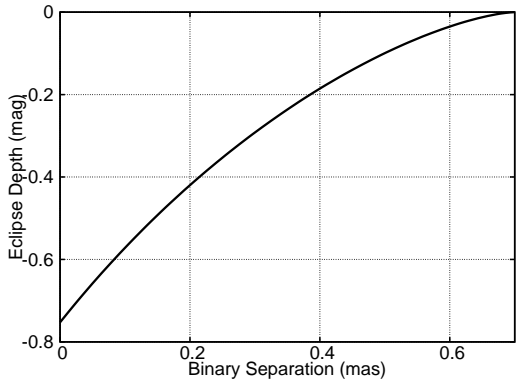}
\caption[Eclipse Depth]
{ \label{fig::EclipseDepth}
Depth of eclipse versus projected binary separation; the shallowest simulation (0.55 mas) predicts an eclipse depth of 0.06 mag.  However, only 0.084 percent of trials (84/100,000) predicted a closest approach of 0.5 mas or more, the point at which the eclipse depth exceeds 0.1 mag.
}
\end{figure}

Figure \ref{fig::EclipseDepth} converts projected sky separation to 
approximate eclipse depth in magnitudes.

\section{Observing Program}

Updated astrometry from the current epoch would greatly improve the eclipse preductions.  For example, a single measurement with mas astrometric precision from early December 2014, when the projected separation is 36 milli-arcseconds, would improve the timing prediction by a factor of 3.  This should be routinely possible with long baseline optical interferometers such as the CHARA array \citep{CHARAEXISTS} or Navy Precision Optical Interferometer \citep[NPOI, ][]{arm98}.  An updated eclipse prediction will be made when these measurements are completed.

\begin{figure}[!ht]
\epsscale{1.1}
\includegraphics[angle=270, width=3.3in]{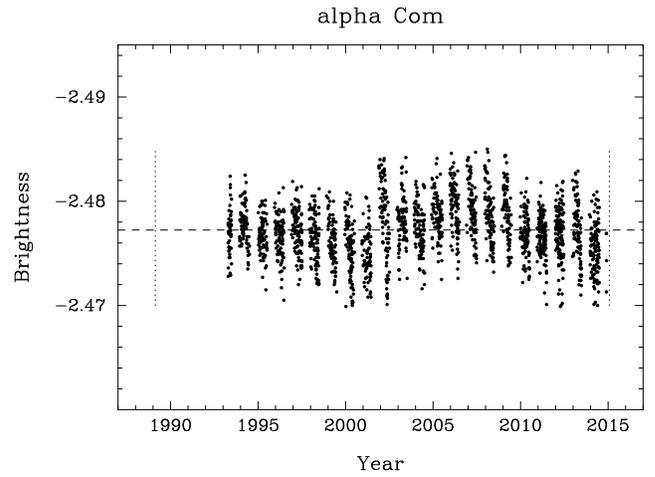}
\caption[HD 114378 ($\alpha$ Com) Photometry 1]
{ \label{fig::114378_phot1}
22 Years of photometry of $\alpha$ Com from the Fairborn T4 APT, relative to the comparison star HD 114520.  The system is slightly variable on decade timescales.  The vertical dotted lines indicated the predicted times of eclipses.
}
\end{figure}

\begin{figure}[!ht]
\epsscale{1.1}
\includegraphics[angle=270, width=3.3in]{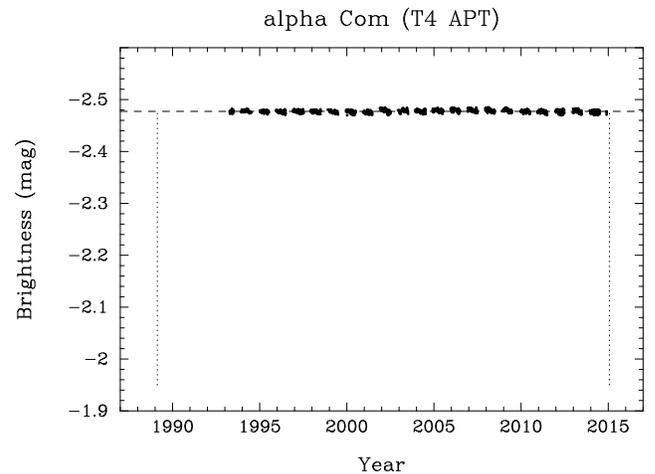}
\caption[HD 114378 ($\alpha$ Com) Photometry 2]
{ \label{fig::114378_phot2}
Fairborn T4 photometry scaled to a partial eclipse depth of 0.5 mag.  Observations began shortly after the previous eclipse and will continue through the upcoming eclipse, pending weather.  Additional observatories are needed to avoid weather concerns.
}
\end{figure}

GWH has observed $\alpha$ Com for 22 years using the Automated Photometric Telescopes at Fairborn Observatory, see Figures \ref{fig::114378_phot1} and \ref{fig::114378_phot2}.  We recommend using the comparison stars HD 114520 (HIP 64312, $V=6.82$, $B-V=0.46$, F2 II) and HD 113848 (39 Com, HR 4946, HIP 63948, $V=5.99$, $B-V=0.39$, F4 V).  Both comparison stars are constant to one millimag (0.001 mag) or so on night-to-night and year-to-year timescales; the standard deviation between comparison stars for the complete 22 year dataset with over 2000 observations is only 0.0018 mag.  $\alpha$ Com itself is slightly variable, ranging over a few mmag from night to night and from year to year (standard deviation 0.0026 mag).  No period has been determined.

Several campaigns have begun to observe the $\alpha$ Com eclipse.  It is particularly important that observations begin well before and continue well after the anticipated eclipse, as this could offer unique opportunities for detecting companion objects such as belts, rings, and possibly even planets.  The AAVSO has organized an observing page at http://www.aavso.org/observing-campaign-alf-Com and an amateur group has organized much background information at http://millimagjournal.wordpress.com/alpha-comae-berenices/ (in French).

\acknowledgements 
The predicted eclipse is near 23 January 2015, MWM's sister's birthday.  If it falls on that day, please refer to it as ``Jessica's Eclipse''.

\bibliography{main}
\bibliographystyle{apj}

\end{document}